# Field of the Magnetic Monopole

A. R. Hadjesfandiari

*Department of Mechanical and Aerospace Engineering*

*State University of New York at Buffalo*

*Buffalo, NY 14260 USA*

ah@eng.buffalo.edu

This paper shows that based upon the Helmholtz decomposition theorem the field of a stationary magnetic monopole, assuming it exists, cannot be represented by a vector potential. Persisting to use vector potential in monopole representation violates fundamentals of mathematics. The importance of this finding is that the vector potential representation was crucial to the original prediction of the quantized value for a magnetic charge.

## §1. Introduction

Quantum mechanics does not require magnetic monopoles to exist. However, Dirac[1] demonstrated that quantum mechanics apparently predicts that a magnetic charge, if it is ever found in nature, must be quantized in units of $\hbar c/2e$ where $e$ is the absolute electron charge value. In this derivation he uses a vector potential to represent the magnetic monopole field. Interestingly, this result has been considered one of the most remarkable predictions of quantum mechanics, which has yet to be verified experimentally.[2] Dirac[3] also developed a general dynamical theory of the magnetic monopole reconfirming and extending his original results. Most authors, even today, present the magnetic monopole quantization based on the first derivation, which Dirac considered incomplete.[2,4,5] Schwinger[6] confirmed this relation using an alternative approach, but with a factor of 2 instead of ½ (i.e., $2\hbar c/e$) and has drawn some speculative conclusions from it. In this paper we investigate the representation of magnetic monopoles in classical electromagnetics, which is a fundamental step in the derivation of Dirac.

We first present several informal arguments based on the concept of continuity and differentiability regarding the Dirac vector potential. We also demonstrate that the Dirac vector potential is actually the vector potential representing the field of a semi-infinite thin solenoid or magnet. After this we rigorously prove that a magnetic monopole field in a static case can only be represented by a scalar potential, not a vector potential as used by Dirac.[1] We show this is simply a direct result of the Helmholtz decomposition theorem. It is seen that there is no vector potential left for Wu and Yang to develop their patching method.[7] Dirac used Stokes' theorem to justify the existence of the arbitrary singular line in the vector potential, representing magnetic monopole. We demonstrate that his reasoning is not consistent with fundamentals of mathematics. We also see that the general dynamical theory of the magnetic monopole[3] still suffers from this inconsistency inherited from Dirac's first paper. Of course, these arguments do not prove or disprove the Schwinger result, but they explain the reason for the discrepancy in the predicted quantized values for the magnetic monopole.

## §2. Magnetic monopole field representation

Suppose, at the origin, there is a point magnetic monopole of strength $q_m$ analogous to a point electric charge. Therefore in Gaussian units

$$\nabla \cdot \mathbf{B} = 4\pi q_m \delta^{(3)}(\mathbf{x}) \tag{1}$$

and the static magnetic field is then given by

$$\mathbf{B} = \frac{q_m}{r^2}\hat{\mathbf{r}} \tag{2}$$

Dirac[1] assumes the concept of vector potential still holds and uses the vector potential in spherical coordinates $(r,\theta,\varphi)$ as

$$\mathbf{A}_1 = \frac{q_m(1-\cos\theta)}{r\sin\theta}\hat{\boldsymbol{\varphi}} = \frac{q_m}{r}\tan\frac{\theta}{2}\hat{\boldsymbol{\varphi}} \quad \theta \neq \pi \tag{3}$$

which apparently satisfies

$$\mathbf{B}_1 = \nabla \times \mathbf{A}_1 = \frac{q_m}{r^2}\hat{\mathbf{r}} \tag{4}$$

This vector potential has a line of singularity along the negative $z$-axis characterized by $\theta = \pi$ called a Dirac string. This line is completely arbitrary and can be taken on any line passing through the origin. But we remember that mathematically, $\mathbf{B}_1$ should be written

$$\mathbf{B}_1 = \nabla \times \mathbf{A}_1 = \frac{q_m}{r^2}\hat{\mathbf{r}} \quad \theta \neq \pi \tag{5}$$

Therefore $\mathbf{A}_1$ and its curl $\mathbf{B}_1$ are not defined on the negative $z$-axis, although $\mathbf{B}_1$ has a limit and a removable singularity equal to the expected $\mathbf{B}$ field. We should also remember that the vector potential $\mathbf{A}$, in a sense, is the integral of the magnetic field $\mathbf{B}$. The magnetic field $\mathbf{B}$ of a monopole is only singular at the origin, but $\mathbf{A}_1$ and $\mathbf{B}_1$ are not defined at the origin and on the negative $z$-axis. An acceptable vector potential $\mathbf{A}$ can have singularity only at the origin. Mathematically the vector fields $\mathbf{B}_1$ and $\mathbf{B}$ are not equivalent, therefore

$$\mathbf{B}_1(\mathbf{x}) \neq \mathbf{B}(\mathbf{x}) \tag{6}$$

A simple analogy from elementary calculus illustrates this point. Let us consider the one variable function $f(s)$ with the derivative given by $f'(s) = 1$ everywhere. By assuming $f(-1) = -1$, we know the function

$$f(s) = s \tag{7}$$

is the desired function. On the other hand, the function

$$f_1(s) = \begin{cases} s & s \leq 0 \\ s+2 & s > 0 \end{cases} \tag{8}$$

also might work. However, there is a discontinuity at $s = 0$ for $f_1(s)$. The derivative of $f_1(s)$ is

$$f_1'(s) = 1 \quad s \neq 0 \tag{9}$$

and is not defined at $s = 0$ although its limit exists. This argument invalidates the discontinuous function $f_1(s)$ in (8) as the integral function. Note the similarity to the situation discussed above for the magnetic monopole. It is incorrect to use the vector potential $\mathbf{A}_1$ in (3) to represent magnetic field of monopole. Also one expects that the isotropic spherical magnetic field in (2) requires an isotropic spherical potential representation. However, the vector potential $\mathbf{A}_1$ in (3) does not posses this symmetry.

The non-physical view concerning the vector potential $\mathbf{A}$ in classical electromagnetic theory is the likely reason for using (3) to represent the magnetic monopole field. In classical electromagnetism, the vector potential has not been given any physical meaning and is considered to be only a mathematical device to simplify the governing equations.

It is interesting to note that by considering $\mathbf{A}$ a non-physical entity, Wu and Yang[7] avoided the singular Dirac string by constructing a pair of potentials

$$\mathbf{A}_1^{(I)} = \frac{q_m(1-\cos\theta)}{r\sin\theta}\hat{\boldsymbol{\varphi}} \qquad \theta < \pi - \varepsilon \tag{10}$$

$$\mathbf{A}_1^{(II)} = -\frac{q_m(1+\cos\theta)}{r\sin\theta}\hat{\boldsymbol{\varphi}} \qquad \theta > \varepsilon \tag{11}$$

such that the potential $\mathbf{A}_1^{(I)}$ can be used everywhere except inside the cone defined by $\theta = \pi - \varepsilon$ around the negative z-axis; likewise the potential $\mathbf{A}_1^{(II)}$ can be used everywhere except inside the cone $\theta = \varepsilon$ around the positive z-axis. Together they apparently lead to the expected expression for $\mathbf{B}$. It is seen that in the overlap region $\varepsilon < \theta < \pi - \varepsilon$ there is not a unique potential vector and

$$\mathbf{A}_1^{(II)} - \mathbf{A}_1^{(I)} = \frac{-2q_m}{r\sin\theta}\hat{\boldsymbol{\varphi}}, \qquad \varepsilon < \theta < \pi - \varepsilon \tag{12}$$

All of this has been justified by arguing that the vector potential is just a device for obtaining $\mathbf{B}$, and we need not insist on having a single expression for $\mathbf{A}$ valid everywhere.[2,7] Our mathematical objection is the same as before. Even by accepting a double valued vector potential, still its curl is not defined on the lines $\theta = \varepsilon$ and $\theta = \pi - \varepsilon$.

We give a counter-example showing that $\mathbf{A}$ cannot be manipulated in an arbitrary manner as Wu and Yang have assumed. The magnetic field of an infinite cylindrical solenoid directed along the z-direction with radius $R$ is given in cylindrical coordinates $(\rho,\varphi,z)$ as[8]

$$\mathbf{B} = \begin{cases} B_0 \hat{\mathbf{z}} & \rho \leq R \\ \mathbf{0} & \rho > R \end{cases} \tag{13}$$

where $B_0$ is the uniform magnetic field inside the solenoid. The vector potential used in the Aharanov-Bohm effect[9] is given by

$$\mathbf{A} = \begin{cases} \dfrac{1}{2} B_0 \rho \hat{\boldsymbol{\varphi}} & \rho \leq R \\ \dfrac{1}{2} B_0 \left(R^2/\rho\right) \hat{\boldsymbol{\varphi}} & \rho > R \end{cases} \tag{14}$$

in cylindrical coordinates. If Wu and Yang are correct, then we should be able to use the vector potential

$$\mathbf{A}_1 = \begin{cases} \dfrac{1}{2} B_0 \rho \hat{\boldsymbol{\varphi}} & \rho \leq R \\ \mathbf{0} & \rho > R \end{cases} \tag{15}$$

which obviously satisfies

$$\mathbf{B} = \nabla \times \mathbf{A}_1 = \begin{cases} B_0 \hat{\mathbf{z}} & \rho \leq R \\ \mathbf{0} & \rho > R \end{cases} \tag{16}$$

However, it is seen that $\mathbf{A}_1$ cannot predict the Aharonov-Bohm effect, because it does not generate the required circulation around the solenoid. It should be noticed that the vector potential given by (14) is the solution to the Poisson equation

$$\nabla^2 \mathbf{A}(\mathbf{x}) = -\frac{4\pi}{c} \mathbf{J}(\mathbf{x}) \tag{17}$$

governing the current generating magnetostatics, where $\mathbf{J}$ is the electric current density.[8] Its general solution is given by

$$\mathbf{A}(\mathbf{x}) = \frac{1}{c} \int \frac{\mathbf{J}(\mathbf{x}')}{|\mathbf{x} - \mathbf{x}'|} d^3 \mathbf{x}' \tag{18}$$

We can add an arbitrary smooth curl free vector which is the manifestation of gauge freedom. From the way Dirac and his followers[1),2),4),5),7)] use the vector potential (3), the impression is that (17) should not be the governing equation for magnetic monopole vector potential. However, we will demonstrate that the Dirac vector potential (3) is really consistent with equation (17) but for a different physical problem.

Before continuing the discussion of the magnetic monopole field, it should be emphasized that there is nothing wrong with having a line of singularity to represent a consistent physical phenomenon. For example, the magnetic field **B** of the long straight wire along the $z$ direction carrying a current $I$ is given in cylindrical coordinates $(\rho, \varphi, z)$ as[8]

$$\mathbf{B} = \frac{2}{c}\frac{I}{\rho}\hat{\boldsymbol{\varphi}} \quad \rho \neq 0 \tag{19}$$

It is seen that the corresponding vector potential is

$$\mathbf{A} = -\frac{2}{c} I \ln \rho \hat{\mathbf{z}} \quad \rho \neq 0 \tag{20}$$

Both vectors **A** and **B** are not defined on the $z$-axis, which in this case corresponds uniquely to the longitudinal axis of the wire. Consequently, this vector potential is acceptable for representing the magnetic field.

As we mentioned the Dirac vector potential in (3) can be used to represent the magnetic field of a physical phenomenon which is consistent with the traditional magnetostatic and satisfies (17) as follows.

*2.1. Field of a long, thin solenoid or magnet*
The Dirac vector potential

$$\mathbf{A}_1 = \frac{q_m(1-\cos\theta)}{r\sin\theta}\hat{\boldsymbol{\varphi}} = \frac{q_m}{r}\tan\frac{\theta}{2}\hat{\boldsymbol{\varphi}} \quad \theta \neq \pi \tag{3}$$

satisfies the Poisson equation

$$\nabla^2 \mathbf{A} = -\frac{4\pi}{c}\mathbf{J}(\mathbf{x}) \tag{17}$$

and can be used to represent the field of a semi-infinite thin solenoid or magnet (very thin means that the cross section approaches to zero). It seems this representation perhaps inspired the initial idea about magnetic monopoles. We demonstrate this fact in details.

For a pure (point-like) magnetic dipole with magnetic moment **m** at the origin, the vector potential is given by[8)]

$$\mathbf{A} = \frac{1}{c}\frac{\mathbf{m}\times\hat{\mathbf{r}}}{r^2} \quad (21)$$

which is consistent with the fundamental Poisson equation (17) for current generating magnetostatics. By letting **m** align in the z-direction, the potential at the point P with spherical coordinates $(r,\theta,\varphi)$ is given by

$$\mathbf{A} = \frac{1}{c}\frac{m\sin\theta}{r^2}\hat{\boldsymbol{\varphi}} \quad (22)$$

and hence

$$\mathbf{B} = \nabla\times\mathbf{A} = \frac{1}{c}\frac{m\sin\theta}{r^3}(2\cos\theta\hat{\mathbf{r}}+\sin\theta\hat{\boldsymbol{\theta}}) \quad (23)$$

Surprisingly, this is identical in structure to the field of an electric dipole. For a pure (point-like) electric dipole with electric dipole moment $\mathbf{p}_e$ at the origin, the scalar electric potential is[8)]

$$\phi_E = \frac{\mathbf{p}_e\cdot\hat{\mathbf{r}}}{r^2} \quad (24)$$

If $\mathbf{p}_e$ is directed in the z-direction, the potential at the point $(r,\theta,\varphi)$ is

$$\phi_E = \frac{p_e\cos\theta}{r^2}\hat{\boldsymbol{\varphi}} \quad (25)$$

and hence

$$\mathbf{E} = -\nabla\phi_E = \frac{p_e\sin\theta}{r^3}(2\cos\theta\hat{\mathbf{r}}+\sin\theta\hat{\boldsymbol{\theta}}) \quad (26)$$

Although the fields **B** and **E** in (23) and (26) are very similar, scalar potential is not used for representing the magnetic dipole field.

For a very thin solenoid or magnet with the distribution of magnet magnetic moment per unit length $\mathbf{M}(\mathbf{x}')$ along its length, the vector potential for point **x** not along the solenoid (magnet) is

$$\mathbf{A}(\mathbf{x}) = \frac{1}{c}\int \frac{\mathbf{M}(\mathbf{x}')\times(\mathbf{x}-\mathbf{x}')}{(\mathbf{x}-\mathbf{x}')^3} dl(\mathbf{x}') \qquad (27)$$

Consider a very thin solenoid (magnet) with length $L$ with uniform magnetic moment per unit length $M$, which is placed along the z-axis as shown in Figure 1. Therefore the vector potential is

$$\mathbf{A}(\mathbf{x}) = \frac{M}{c}\left(\int_{-L}^{0} \frac{\sin\theta'}{r'^2} dz'\right)\hat{\boldsymbol{\varphi}} \qquad \mathbf{x}\notin OA \qquad (28)$$

By using the sine and cosine laws in Figure 2 we obtain

$$\frac{r}{\sin\theta'} = \frac{r'}{\sin\theta} \qquad (29)$$

$$r'^2 = r^2 + z'^2 - 2rz'\cos\theta \qquad (30)$$

We can express all variable in terms variable $z'$ as

$$\mathbf{A}(\mathbf{x}) = \frac{M}{c} r\sin\theta \left[\int_{-L}^{0} \frac{dz'}{(r^2+z'^2-2rz'\cos\theta)^{\frac{3}{2}}}\right]\hat{\boldsymbol{\varphi}} \qquad \mathbf{x}\notin OA \qquad (31)$$

After integration we obtain

$$\mathbf{A}(\mathbf{x}) = \frac{1}{c}\frac{M}{r\sin\theta}(\cos\theta_2 - \cos\theta)\hat{\boldsymbol{\varphi}} \qquad \mathbf{x}\notin OA \qquad (32)$$

It can be easily shown that

$$\mathbf{B}(\mathbf{x}) = \nabla\times\mathbf{A}(\mathbf{x}) = \frac{1}{c}\frac{M}{r^2}\hat{\mathbf{r}} - \frac{1}{c}\frac{M}{r_2^2}\hat{\mathbf{r}}_2 \qquad \mathbf{x}\notin OA \qquad (33)$$

Actually this equation is the reason behind the assumption of a possible existence of magnetic monopoles. It looks there are two point magnetic poles with charges $q_m$, and $-q_m$ at points O and A, where

$$q_m = \frac{M}{c} \qquad (34)$$

producing the vector potential

$$\mathbf{A}(\mathbf{x}) = \frac{q_m}{r\sin\theta}(\cos\theta_2 - \cos\theta)\hat{\boldsymbol{\varphi}} \qquad \mathbf{x} \notin OA \qquad (35)$$

And magnetic field

$$\mathbf{B}(\mathbf{x}) = \frac{q_m}{r^2}\hat{\mathbf{r}} - \frac{q_m}{r_2^2}\hat{\mathbf{r}}_2 \qquad \mathbf{x} \notin OA \qquad (36)$$

This is similar to the electric dipole consisting two charges $q$ and $-q$ at the same points. The corresponding scalar electric field is

$$\phi_E(\mathbf{x}) = \frac{q}{r} - \frac{q}{r_2} \qquad (37)$$

and hence the electric field is

$$\mathbf{E}(\mathbf{x}) = -\nabla\phi_E(\mathbf{x}) = \frac{q}{r^2}\hat{\mathbf{r}} - \frac{q}{r_2^2}\hat{\mathbf{r}}_2 \qquad (38)$$

But we should remember that the potential vector $\mathbf{A}$ and magnetic field $\mathbf{B}$ in (35) and (36) are not defined along the line OA representing the solenoid or magnet, but the scalar electric potential $\phi_E$ and electric field $\mathbf{E}$ in (37) and (38) are defined on this line except at the points O and A. Therefore, it is clear that although the electric field $\mathbf{E}$ in (38) and magnetic field $\mathbf{B}$ in (36) might look similar, they are not mathematically equivalent. The appearance of different potentials is the manifestation of this fact.

Interestingly, it is seen that for the case of a semi-infinite solenoid (magnet) where $L \to \infty$ ($\theta_2 \to 0$), we have

$$\mathbf{A}(\mathbf{x}) = \frac{q_m(1-\cos\theta)}{r\sin\theta}\hat{\boldsymbol{\varphi}} \qquad \theta \neq \pi \qquad (39)$$

and

$$\mathbf{B}(\mathbf{x}) = \frac{q_m}{r^2}\hat{\mathbf{r}} \qquad \theta \neq \pi \qquad (40)$$

which are exactly the same as the vector fields (3) and (4) used by Dirac to represent a monopole field. As before $\mathbf{A}(\mathbf{x})$ and $\mathbf{B}(\mathbf{x})$ are not defined along the negative $z$-axis. This is because the magnet or solenoid is on this axis which represents the distribution of the source current. However, a real monopole should generate an isotropic spherical field

$$\mathbf{B}(\mathbf{x}) = \frac{q_m}{r^2}\hat{\mathbf{r}} \tag{2}$$

which is not equivalent to (40). Trying to use the results for a semi-infinite solenoid to represent the field of a magnetic monopole obviously is not correct. The line of singularity has a physical meaning for the solenoid (magnet), but it has been created for monopole due to using vector potential (3). Dirac justifies this by incorrectly using the Stokes theorem which will be inspected later in this paper. He and his followers are actually replacing a semi-infinite solenoid with a point monopole. It seems nobody can argue that the mentioned vector potential is correct for representing both magnetic monopole and thin semi-infinite solenoid. It does not seem correct to consider a non Euclidean geometry or recourse to bundle theory for the case of magnetic monopole as Wu and Yang have done.[7), 10)]

As we have suspected, representing a magnetic monopole field by a vector potential might not be allowed. Must we abandon the use of the vector potential? We should recall that the potential vector $\mathbf{A}$ is defined as

$$\mathbf{B} = \nabla \times \mathbf{A} \tag{41}$$

by considering the condition

$$\nabla \cdot \mathbf{B} = 0 \tag{42}$$

everywhere in space.[8)]

By assuming the existence of a point monopole satisfying

$$\nabla \cdot \mathbf{B} = 4\pi q_m \delta^{(3)}(\mathbf{x}) \tag{1}$$

we can no longer use the potential vector $\mathbf{A}$ exclusively to obtain $\mathbf{B}$, because now $\mathbf{B}$ is not solenoidal. This is the result of the Helmholtz decomposition theorem for vector fields,[11)] which reads:

*Helmholtz decomposition (resolution) theorem:* If the divergence and curl of a vector function **B(x)** are specified as

$$\nabla \cdot \mathbf{B} = \alpha(\mathbf{x}) \tag{43}$$

$$\nabla \times \mathbf{B} = \boldsymbol{\beta}(\mathbf{x}) \tag{44}$$

(for consistency, **β** must be divergence-less,

$$\nabla \cdot \boldsymbol{\beta} = 0 \tag{45}$$

because the divergence of a curl is always zero), and if both $\alpha(\mathbf{x})$ and $\boldsymbol{\beta}(\mathbf{x})$ go to zero faster than $1/r^2$ as $r \to \infty$, and if **B(x)** goes to zero as $r \to \infty$, then **B(x)** is given uniquely by

$$\mathbf{B} = -\nabla \phi + \nabla \times \mathbf{A} \tag{46}$$

where

$$\phi(\mathbf{x}) = \frac{1}{4\pi} \int \frac{\alpha(\mathbf{x}')}{|\mathbf{x} - \mathbf{x}'|} d^3\mathbf{x}' \tag{47}$$

and

$$\mathbf{A}(\mathbf{x}) = \frac{1}{4\pi} \int \frac{\boldsymbol{\beta}(\mathbf{x}')}{|\mathbf{x} - \mathbf{x}'|} d^3\mathbf{x}' \tag{48}$$

It should be emphasized that $\alpha(\mathbf{x})$ and $\boldsymbol{\beta}(\mathbf{x})$ can be distribution functions which require interpreting integrals (47) and (48) in a distributional sense. In our discussion for the magnetic monopole, we have

$$\alpha(\mathbf{x}) = \nabla \cdot \mathbf{B} = 4\pi q_m \delta^{(3)}(\mathbf{x}) \tag{49}$$

and

$$\boldsymbol{\beta}(\mathbf{x}) = \nabla \times \mathbf{B} = 0 \tag{50}$$

where $\alpha(\mathbf{x})$ and $\boldsymbol{\beta}(\mathbf{x})$ are distribution or generalized functions. It is seen that the assumed magnetic field of a monopole given by

$$\mathbf{B} = \frac{q_m}{r^2} \hat{\mathbf{r}} \tag{2}$$

is curl free even at the origin in a distributional sense which justifies (50). Therefore, all of the conditions of the theorem are satisfied. Then from (47) and (48), we find

$$\phi(\mathbf{x}) = \frac{q_m}{r} \tag{51}$$

and

$$\mathbf{A}(\mathbf{x}) = 0 \tag{52}$$

respectively. We only have the scalar potential $\phi$ to represent the vector field $\mathbf{B}$ through

$$\mathbf{B} = -\nabla \phi \tag{53}$$

Therefore

$$\mathbf{B} = \frac{q_m}{r^2} \hat{\mathbf{r}} \tag{2}$$

with no arbitrary line of singularity in $\phi$ and its gradient.

It should be mentioned that if we release the constraint forcing $\mathbf{B}(\mathbf{x})$ to zero as $r \to \infty$, we can add an arbitrary constant to the potential $\phi$ and an arbitrary regular curl free vector to $\mathbf{A}(\mathbf{x})$. Of course this does not change anything in our conclusion.

We have for magnetostatics something similar to electrostatics with the expected spherical symmetry. It is seen that the vector potential $\mathbf{A}(\mathbf{x})$ does not exist for magnetic monopole. Based on the Helmholtz theorem, the magnetic monopole, if it exists, must be similar to an electric charge. Actually the relation (53) was used in the beginning to obtain (2) by indirect analogy with electrostatics. The representation of the field $\mathbf{B}$ by only the vector potential $\mathbf{A}(\mathbf{x})$ violates the Helmholtz decomposition theorem. The existence of a line of singularity in $\mathbf{A}(\mathbf{x})$ is just the manifestation of that violation.

Attempting to represent a stationary magnetic monopole field by a vector potential is exactly similar to representing a stationary point electric charge by a vector potential. A stationary electric charge field is irrotational and is represented only by a scalar potential. This is the direct result of the Helmholtz theorem which also requires a scalar potential representation for the magnetostatic field of a monopole.

In general every vector field, such as the magnetic field, can be represented by

$$\mathbf{B} = -\nabla \phi + \nabla \times \mathbf{A} \tag{46}$$

The requirement of

$$\nabla \bullet \mathbf{B} = 0 \tag{42}$$

which is the conventional relation in electrodynamics leaves us with

$$\mathbf{B} = \nabla \times \mathbf{A} \tag{41}$$

Otherwise the general equation is (46), not (41).

In some presentations when there is a magnetic charge distribution $\rho_m(\mathbf{x})$ such that

$$\nabla \bullet \mathbf{B} = 4\pi \rho_m \tag{54}$$

authors admit that using vector potential $\mathbf{A}$ is not correct.[4)] However, $\mathbf{A}$ is still used for a magnetic monopole at the origin by considering

$$\nabla \bullet \mathbf{B} = 0 \quad \text{for} \quad \mathbf{x} \neq \mathbf{0} \tag{55}$$

But this relation is not complete because it does not consider the magnetic charge distribution. The magnetic charge distribution is defined by the generalized function

$$\rho_m = q_m \delta^{(3)}(\mathbf{x}) \tag{56}$$

where $\delta^{(3)}(\mathbf{x})$ is (ironically) the Dirac delta function. By using this generalized function, (54) becomes

$$\nabla \bullet \mathbf{B} = 4\pi q_m \delta^{(3)}(\mathbf{x}) \tag{1}$$

It is this equation that has the complete mathematical meaning and therefore enables us to consider the magnetic monopole correctly, as has been presented here. Apparently, these authors do not realize the consequences of representing the source point by a distribution function and the applicability of the Helmholtz decomposition theorem. Thus, they implicitly assume that one point does not cause any problem and then utilize the vector potential.

It seems that adopting a more mathematical viewpoint is helpful. We notice that we have here an elliptic singular exterior boundary value problem. The bounded boundary can be considered to be the surface of a sphere that has approached to a point. This provides a point singularity, but the total assumed magnetic charge is constant. Distribution theory enables us to use the Dirac delta function for representing the source. From the theory of elliptic boundary value problems, we know the solution is analytic in the domain. Therefore, the derivatives and integrals have to be analytic. This is the complement of what was presented in the beginning of this section as speculations. The Dirac vector potential clearly is not analytic along the line of singularity. Similarly, the composite Wu-Yang solution[7),10)] also is not analytic throughout the domain. Interestingly, their double-valued vector potential is not acceptable for the physical semi-infinite thin solenoid, because the line of singularity is a reality. Based on the Helmholtz decomposition theorem, no vector potential exists to represent the magnetic monopole field.

It is interesting to note that Dirac[1)] argues that by choosing a vector potential representing the magnetic monopole, it must have a singularity line which does not need to be even straight. Dirac uses this argument as a fundamental step in his more extended version.[3)] We demonstrate this argument following modern versions.[4),5)]

First we need to present Stokes' theorem which transforms contour integrals to surface integrals and vice versa.[11)]

*Stokes' theorem:* Let $\mathbf{V}(\mathbf{x})$ be a single-valued vector function which is continuous and differentiable everywhere in a surface $S$ and on a simple closed contour $C$ that bounds $S$. Then

$$\oint_C \mathbf{V} \bullet d\mathbf{l} = \int_S (\nabla \times \mathbf{V}) \bullet d\mathbf{S} \qquad (57)$$

This states that the line integral of $\mathbf{V}(\mathbf{x})$ over the contour equals the surface integral of $\nabla \times \mathbf{V}$ over a surface $S$ bounded by $C$.

It is also seen that if a continuous vector field $\boldsymbol{\omega}(\mathbf{x})$ is given on the surface $S$ it is possible to obtain a smooth vector field $\mathbf{V}$ such that

$$\int_S \boldsymbol{\omega} \bullet d\mathbf{S} = \oint_C \mathbf{V} \bullet d\mathbf{l} \qquad (58)$$

where

$$\boldsymbol{\omega} = \nabla \times \mathbf{V} \quad \text{on the surface } S \qquad (59)$$

For the magnetic monopole at the origin, consider a closed loop $C$ at fixed $r, \theta$ with $\varphi$ ranging from 0 to $2\pi$. The total flux $\Phi(r,\theta)$ passing through the spherical cap surface $S$ defined by this particular $r$, $\theta$ and bounded from below to the closed loop $C$ is

$$\Phi(r,\theta) = \int_S \mathbf{B} \bullet d\mathbf{S} = 2\pi q_m (1 - \cos\theta) \tag{60}$$

Based on Stokes' theorem this flux can be written as a contour integral of vector $\mathbf{A}'$ on the loop $C$ as

$$\Phi(r,\theta) = \oint_C \mathbf{A}' \bullet d\mathbf{l} \tag{61}$$

As $\theta$ is varied the flux through the cap varies. For $\theta \to 0$ the loop shrinks to a point and the flux passing through the cap approaches zero

$$\Phi(r,0) = 0 \tag{62}$$

As the loop is lowered over the sphere, the cap encloses more and more flux until, eventually, at $\theta \to \pi$ we should have

$$\Phi(r,\pi) = 4\pi q_m \tag{63}$$

which is the total flux passing through the spherical surface enclosing the point charge. However, as $\theta \to \pi$ the loop has again shrunk to a point so the requirement that $\Phi(r,\pi)$ be finite entails from (18) that $\mathbf{A}'$ is singular at $\theta = \pi$. This argument holds for all spheres of all possible radii, so it follows that $\mathbf{A}'$ is singular along the entire negative $z$-axis. It is clear that by a suitable choice of coordinates the string may be chosen to be along any direction, and, in fact, need not be straight, but must be continuous. We should be aware that only

$$\mathbf{B} = \nabla \times \mathbf{A}' \quad \text{on the surface } S \tag{64}$$

For having

$$\mathbf{B} = \nabla \times \mathbf{A}' \tag{65}$$

to be correct except at the origin, it is necessary that Stokes' theorem be correct for every arbitrary contour and surface cap. The surface cannot be arbitrary because it must not intersect the singular string line. Therefore, we can see that the familiar vector

$$\mathbf{A}' = \mathbf{A}_1 = \frac{q_m (1 - \cos\theta)}{r \sin\theta} \hat{\boldsymbol{\varphi}} \quad \theta \neq \pi \tag{66}$$

satisfies the Stokes theorem (61) to calculate the flux $\Phi(r,\theta)$ only on the mentioned spherical cap which does not intersect with the string line. It is seen that $\mathbf{A}'$ is just a mathematical tool to transform a contour integral to a surface integral without having any physical meaning. Because of this we have called $\mathbf{A}'$ a vector field not a vector potential. As it was shown above, the singularity string in $\mathbf{A}'$ depends on the original closed loop $C$.

This argument may be clarified further by considering the electric flux of a point electric charge $q$. The electric field

$$\mathbf{E} = \frac{q}{r^2}\hat{\mathbf{r}} \tag{67}$$

satisfies

$$\nabla \bullet \mathbf{E} = 4\pi q \delta^{(3)}(\mathbf{x}) \tag{68}$$

The electric flux passing through the mentioned cap is

$$\Phi_E(r,\theta) = \int_S \mathbf{E} \bullet d\mathbf{S} = 2\pi q(1-\cos\theta) \tag{69}$$

We can also consider the vector field $\mathbf{A}'_E$ such that

$$\mathbf{E} = \nabla \times \mathbf{A}'_E \quad \text{on the surface } S \tag{70}$$

By using an argument analogous to that employed by Dirac,[1] we can see $\mathbf{A}'_E$ has an arbitrary line of singularity. This vector field obviously can be taken as

$$\mathbf{A}'_E = \frac{q(1-\cos\theta)}{r\sin\theta}\hat{\boldsymbol{\varphi}} \quad \theta \neq \pi \tag{71}$$

which satisfies

$$\Phi_E(r,\theta) = \oint_C \mathbf{A}'_E \bullet d\mathbf{l} \tag{72}$$

But using $\mathbf{A}'_E$ to represent the electric field everywhere is inconsistent with the electrostatic theory which uses only a scalar potential to represent the electric field of a point electric charge at rest. We see that if a magnetic monopole exists, then magnetostatics should be similar to electrostatics, and must be represented by scalar potential. The vector fields $\mathbf{A}'$ and $\mathbf{A}'_E$ are only suitable to transform integrals from contour to surface and vice versa, although doing so is a redundant action.

As it was mentioned earlier the problem originates in replacing a semi-infinite solenoid or magnet with a point magnetic monopole.

### §3. Quantum mechanics magnetic monopole quantization

As we mentioned previously, the magnetic monopole representation by a potential is very crucial to the quantum mechanics step, which we now briefly discuss. The Dirac derivation[1] is based on the single valuedness of the wave function around the singular string. However, we have seen that this derivation is not consistent with the Helmholtz decomposition theorem.

The classical Hamiltonian for a free charged particle with momentum **p** is

$$H = \frac{1}{2m} p^2 \tag{73}$$

The Schrödinger equation for a free particle is obtained from this classical Hamiltonian by the usual prescription[2]

$$\mathbf{p} \to \frac{\hbar}{i} \nabla \tag{74}$$

$$H \to -\frac{\hbar}{i} \frac{\partial}{\partial t} \tag{75}$$

Letting the operators act on the wave function $\psi$, then the Schrödinger equation for a free particle is

$$-\frac{\hbar^2}{2m} \nabla^2 \psi = -\frac{\hbar}{i} \frac{\partial \psi}{\partial t} \tag{76}$$

It has been shown that for a charged particle in an electromagnetic field we need to use minimal substitutions.[2] This consists of the following replacements

$$\mathbf{p} \to \frac{\hbar}{i} \nabla - \frac{q}{c} \mathbf{A} \tag{77}$$

$$H \to -\left( \frac{\hbar}{i} \frac{\partial}{\partial t} + q\phi \right) \tag{78}$$

The Schrödinger equation then becomes

$$\frac{1}{2m}(\frac{\hbar}{i}\nabla - \frac{q}{c}\mathbf{A})^2\psi + q\phi\psi = -\frac{\hbar}{i}\frac{\partial\psi}{\partial t} \tag{79}$$

For motion of an electron with electric charge $q = -e$ in the field of a magnetic monopole with charge $q_m$, Dirac and his followers[1),2),4,)5,7)] consider the equation

$$\frac{1}{2m}(\frac{\hbar}{i}\nabla - \frac{q}{c}\mathbf{A})^2\psi = -\frac{\hbar}{i}\frac{\partial\psi}{\partial t} \tag{80}$$

with

$$\mathbf{A} = \frac{q_m(1-\cos\theta)}{r\sin\theta}\hat{\boldsymbol{\phi}} \tag{81}$$

From our discussion we can see that this is not correct. The correct equation is

$$-\frac{\hbar^2}{2m}\nabla^2\psi + q\phi\psi = -\frac{\hbar}{i}\frac{\partial\psi}{\partial t} \tag{82}$$

where

$$\phi(\mathbf{x}) = \frac{q_m}{r} \tag{51}$$

It is obvious that this form of the Schrödinger equation cannot produce the original quantization relation for magnetic monopole charges.[1)]

Dirac considered his derivation in 1931, which we explained here, incomplete and developed a general dynamical theory of the magnetic monopole.[3)] However, in this more complicated derivation he retained his original results.

In the usual case, the anti-symmetric electromagnetic strength tensor

$$F^{\alpha\beta} = \begin{pmatrix} 0 & -E_1 & -E_2 & -E_3 \\ E_1 & 0 & -B_3 & B_2 \\ E_2 & B_3 & 0 & -B_1 \\ E_3 & -B_2 & B_1 & 0 \end{pmatrix} \tag{83}$$

is represented by the four-vector potential

$$A^\alpha = (\phi, \mathbf{A}) \tag{84}$$

as

$$F^{\alpha\beta} = \partial^\alpha A^\beta - \partial^\beta A^\alpha \tag{85}$$

Dirac considers this incorrect[3] and adds a new term in the strength tensor as

$$F^{\alpha\beta} = \partial^\alpha A^\beta - \partial^\beta A^\alpha + 4\pi G^{\alpha\beta} \tag{86}$$

with some required condition. Although the addition of the tensor $4\pi G^{\alpha\beta}$ may be viewed in hindsight as an attempt to satisfy the Helmholtz theorem, unfortunately he still considered the existence of the assumed arbitrary string line.[1] As we already explained this string line is not consistent with mathematical rules such as Stokes' theorem, as well as the Helmholtz theorem. In particular, the Helmholtz theorem does not allow us to be so free in considering a singularity line and so on.

As mentioned previously, considering **A** as a device without physical meaning might have been the reason to violate mathematical rules. However, in the quantum mechanical relation (79) potentials $\phi$ and **A** are considered physical. In the explanation of the Aharonov-Bohm effect[9] some researchers have concluded that in quantum mechanics it is **A** rather than **B** that is fundamental as a physical reality. It is more surprising to note that the derivation of the magnetic monopole and the explanation of the Aharonov-Bohm effect are essentially based on the same ideas related to the gauge transformations in quantum mechanics.[2,7,10] We conclude that, for the electromagnetic field of a magnetic monopole, the potentials $\phi$ and **A** should be also physical and certainly must satisfy mathematical rules such as the Helmholtz decomposition theorem. In our discussion we have only used mathematical rules to disqualify the magnetic monopole derivation. However, finding a physical reality for scalar and vector potentials in classical electrodynamics can render an obvious reason for non-physicality of the singularity or discontinuity on an arbitrary line in the **A** field.

From the discussion in this paper, one might speculate that magnetic monopoles do not exist. The lack of finding even one monopole is perhaps good support for this speculation. Why does nature need to have electrostatic-like monopoles? However, this paper is not about disproving the existence of magnetic monopoles. This speculation will be discussed further in a future paper.

## §4. Conclusion

It has been shown that based on the Helmholtz decomposition theorem, the stationary magnetic monopole field can only be represented by a scalar potential similar to electrostatics, not as a vector potential. Considering the vector potential **A** as a non-physical entity, we still cannot use it in representing a magnetic monopole, because of the violation of the Helmholtz decomposition theorem. Vector potential **A** cannot be manipulated as long as it generates magnetic field. Its continuity must be guaranteed. The smoothness of the magnetic strength field **B** everywhere, except the position of monopole, must be satisfied by every integral representation of this field. It was shown that the Dirac vector potential for a magnetic monopole actually is the vector potential representing the field of a semi-infinite thin solenoid or magnet. This vector potential cannot represent the field of two different physical phenomena at the same time. The line of singularity for the solenoid (magnet) is physical and acceptable, but it is not for a point monopole having isotropic spherical symmetry. This fact prevents us to consider a non Euclidean geometry or to use fiber bundle theory for explaining the magnetic monopole.

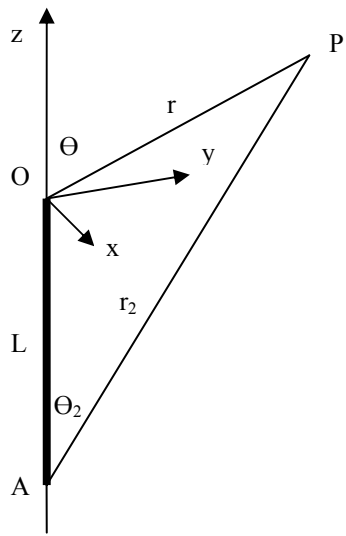

Figure 1

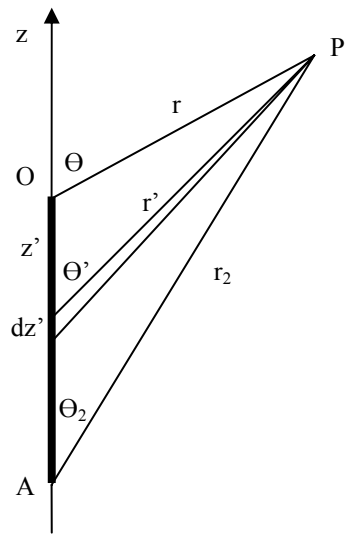

Figure 2